\documentclass[10pt,aps,prc,twocolumn,amsmath,amssymb,floatfix,superscriptaddress]{revtex4-2}
\usepackage{mathptmx}
\DeclareSymbolFont{cmm}{OML}{cmm}{m}{it}
\DeclareMathSymbol{\varrho}{\mathalpha}{cmm}{"25}
\usepackage{CJK}
\usepackage{graphicx}
\usepackage{xcolor}
\usepackage{bm}
\usepackage{booktabs,array}
\usepackage[breaklinks,pdfstartview=FitH,bookmarksnumbered=true,
bookmarksopen=true,colorlinks=true,linkcolor=blue,urlcolor=blue,citecolor=blue]{hyperref}
\DeclareMathAlphabet{\mathcal}{OMS}{cmsy}{m}{n}

\begin{document}
    \title{Pairing connections between localized nuclear regions: Orbital reweighting and thermal quenching}
    \begin{CJK*}{UTF8}{gbsn}
        \author{Huai-Tong~Xue~(薛怀通)}\email[Contact author, ]{htxue@nynu.edu.cn}
        \author{Ning-Ning~Gong~(宫宁宁)}
        \affiliation{College of Physics and Electronic Engineering, Nanyang Normal University, Nanyang 473061, China}
        \author{Xian-Rong~Zhou~(周先荣)}\email[Contact author,]{xrzhou@phy.ecnu.edu.cn}
        \affiliation{School of Physics, East China Normal University, Shanghai 200241, China}
\begin{abstract}
\begin{description}
\item[Background] The nucleon localization function (NLF) characterizes nuclear spatial structure, while the nonlocal anomalous density, or pairing tensor, describes the coherent pair amplitude between two points. Global pairing measures leave the connections between distinct spatial regions unresolved.
\item[Purpose] We determine how selective orbital participation strengthens the pairing connection between localized regions during overall pairing loss, and why heating can instead attenuate the connection while preserving its pattern.
\item[Methods] Axial Hartree--Fock plus BCS calculations combine the NLF with the pairing tensor in shape-constrained $^{20}$Ne, with distinct localization lobes, and octupole-deformed $^{144}$Ba, which provides a reflection-asymmetric background. Complete signed orbital sums separate anomalous-weight and spatial-kernel changes at fixed pairing strength and cutoff within each comparison.
\item[Results] Contraction along a constrained $^{20}$Ne branch preserves identifiable end lobes while reducing the total anomalous norm and increasing the absolute opposite-end integral. The coupled spectrum and gap changes favor a mixed $1d_{5/2}/2s_{1/2}$ component and deplete the anomalous participation of a predominantly $1p_{1/2}$ component. Increased positive strength and weaker destructive interference overcome the net opposing spatial-kernel response; this balance survives mesh refinement. Heating at the reference quadrupole constraint yields compensating thermal and quasiparticle-energy factors, so similar gap reductions produce nearly common attenuation. In $^{144}$Ba, octupole deformation and normal-localization modulation persist after the anomalous sources become negligible.
\item[Conclusions] Regional enhancement depends on how selective anomalous participation changes the complete signed orbital sum, including spatial feedback. Nearly proportional participation changes preserve the pattern when the kernels remain stable. These conditions distinguish stronger absolute connections from larger normalized fractions on an identifiable localization background.

\end{description}
\end{abstract}
\maketitle
\end{CJK*}

\section{Introduction}
Clustering and pairing provide complementary perspectives on nuclear structure. Cluster configurations reveal spatial organization, while pairing describes coherent nucleon pairs, particularly among states near the Fermi surface \cite{Freer2018,Dean2003}. Their spatial properties can be studied with the nucleon localization function (NLF) and the nonlocal anomalous density, or pairing tensor. Both are constructed from the single-particle orbitals, but their different normal and anomalous weights allow them to respond differently to changes of the nuclear state.

The NLF resolves shell and spatial structure beyond the smooth density \cite{Reinhard2011,Zhang2016Localization}. Confinement, deformation and angular-momentum alignment shape its patterns \cite{Ebran2012,Ebran2013,Ebran2014,Li2020Rotation}; constrained dilution and nuclear collisions reveal cluster configurations and transient structures \cite{Girod2013prl,Schuetrumpf2017prc}. Density support remains essential for identifying compact clusters from high-NLF regions \cite{Khan_2022_prc}. This normal-density measure describes short-range same-spin exchange, whereas interaction-induced two-body correlations require additional information \cite{Bulgac2023}.

The pairing tensor $\widetilde\kappa(1,2)$, with position and spin included in each index, describes the coherent anomalous amplitude between two points. Here $p(R,r)$ denotes the normalized strength distribution of its cutoff-regulated spin-singlet component in the pair-center and relative radii $R$ and $r$. Near-Fermi angular momenta and bare or induced interactions shape the spatial pairing field \cite{Sandulescu2005,Pastore2008}; pair and quartet coherence lengths probe distinct correlations \cite{Delion2015}. Low density and weak binding favor compact pair correlations \cite{Matsuo2005Dineutron,Matsuo2006Spatial,Hagino2007Coexistence}, while finite-size orbital structure also produces small surface pair sizes \cite{Pillet2007,Pillet2010,Vinas2010Sizes}. Pillet et al. showed how reduced parity cancellation raises a normalized long-distance peak when the pairing force weakens \cite{Pillet2010}. Lasseri et al. identified the roles of relative anomalous weights and orbital localization in coherence lengths, including cancellation of proportional weight changes in a normalized measure \cite{Lasseri2018}.

These studies connect orbital participation to the anomalous density and to normalized measures such as pair sizes. A remaining question is how the absolute anomalous strength connecting identifiable localization regions responds when the nuclear configuration changes self-consistently. Normalization removes the overall strength, so an increased regional fraction need not imply a stronger connection. We ask which changes of the spectrum, pairing amplitudes and orbital spatial support allow a regional connection to strengthen during overall pairing loss, and which instead preserve its pattern during attenuation.

Deformation and temperature provide two ways to address this question. Deformation changes the single-particle spectrum and orbital spatial support. Heating changes thermal occupations together with the self-consistent gaps and mean fields. Earlier finite-temperature studies relate localization to pairing, shape evolution and dilution \cite{Yuksel2022,Davies2025Thermal}. Comparing the two perturbations tests whether an identifiable localization background accompanies selective redistribution of the anomalous amplitude or a nearly proportional loss of strength. Following the source toward quenching also tests whether normal localization persists when the anomalous connection becomes negligible.

The mixed cluster and deformed mean-field character of $^{20}$Ne \cite{Kimura2004,Mehl2025,Xue2025} motivates comparing constrained configurations with identifiable localization lobes. Projected and configuration-mixed studies connect its intrinsic structures to collective bands and shape coexistence \cite{marevic2018prc,Xue2024prc2}. We compare contraction with heating at the reference quadrupole constraint, allowing the remaining fields and orbitals to respond self-consistently. The established octupole collectivity of $^{144}$Ba \cite{Bucher2016PRL} motivates a second thermal comparison on a reflection-asymmetric background, with freely evolving quadrupole and octupole moments.

We use axial Hartree--Fock plus BCS (HF+BCS) calculations to combine the NLF with the nonlocal anomalous density. In Ne, NLF valleys delimit the end regions; in Ba, the NLF tracks normal structure across a geometrically defined half-space partition. Absolute regional integrals and normalized pairing maps distinguish connection strength from its distribution. Complete signed orbital sums connect the spectrum and gap response to changes in anomalous participation, interference and spatial support. Section~\ref{sec:framework} presents the theoretical framework and calculation settings. Section~\ref{sec:results} develops this orbital explanation for deformation and heating and discusses its implications; Sec.~\ref{sec:conclusions} summarizes the conclusions.

\section{Theoretical framework}\label{sec:framework}
\subsection{HF+BCS calculations and intrinsic configurations}
The axial HF+BCS calculations \cite{Bender2003} use SkyAx \cite{SkyAx2021}, extended for self-consistent finite-temperature pairing and spatial analysis, with the SLy5 functional \cite{Chabanat1998} and density-dependent delta pairing \cite{Bender2000}. The attractive strengths $V_p=1146$ and $V_n=999$ MeV fm$^3$ multiply the density factor $1-(\rho_n+\rho_p)/\rho_c$, with $\rho_c=0.16$ fm$^{-3}$. Their SkI4+BCS calibration to odd--even mass differences used a self-adjusting smooth window \cite{Bender2000}; no refit is made for SLy5 or the present frozen window. The anomalous amplitude has no imposed lower bound. The cold Ne reference supplies a 48-entry proton/neutron cutoff vector, frozen by orbital slot throughout the shape and thermal comparisons. For zero-range pairing, this regulator is part of the interaction prescription \cite{BulgacYu2002,Borycki2006,Lalit2026FiniteRange}.

The Ne states use a 0.4 fm axial and radial mesh, with a 0.3 fm cold comparison retaining the same orbital pool and cutoff. The 78 axial and 39 radial nodes span $-15.6\leq z\leq15.2$ fm and $0\leq\varrho\leq15.2$ fm. Each species contains 24 Kramers-pair representatives in a full-$z$ representation with zero odd-moment targets. The damped quadrupole constraint is
\begin{equation}
 q_2=\frac{1}{A s^2}\int(2z^2-\varrho^2)e^{-(r/R_c)^8}\rho_N(\mathbf r)\,d^3r.
\end{equation}
Here $\varrho$ is the cylindrical radius, $\rho_N=\rho_n+\rho_p$, and $s=1.2A^{1/3}$ fm. The prescribed scale $R_c=8$ fm suppresses the constraint field in the outer box. The elongated reference $D_0$ is a prescribed slice inherited from the initial shape screening, with $q_2(D_0)=3(0.9)/\sqrt{5\pi}=0.68125$ and 0.9 an input deformation proxy. Its density-supported NLF end lobes and appreciable anomalous source permit a comparison of pairing connections between identifiable regions. The more compact target $q_2(D_-)=0.70q_2(D_0)=0.47687$ retains distinguishable end regions under contraction; heating retains the $D_0$ target. These damped moments differ from $\beta_2$. The targets were fixed before the response comparison and specify paired intrinsic slices, not energy minima or laboratory states (Table~\ref{tab:anchors}). SLy5 retains its central $\mathbf J^2$ terms; the full center-of-mass correction is evaluated after variation with the unmodified mean-field kinetic coefficient \cite{Bender2000CM}.

The reflection-asymmetric $^{144}$Ba calculations retain SLy5 and the same pairing strengths, with a variational one-body center-of-mass correction and complete pools of 114 proton and 155 neutron orbitals. Cold states initialized with a small reflection-asymmetric perturbation seed the thermal continuation. The center of mass is constrained while quadrupole and octupole moments evolve freely. The main comparison uses a 0.6 fm mesh with radial and axial box extents of 24 fm, and separate 0.6 and 0.8 MeV continuations from the same cold state. A 19.8 fm sequence at $T=0$, 0.3 and 0.6 MeV provides a finite-volume comparison; a 30 fm cold state tests the interior pattern. Each box retains its own cold cutoff vector (Supplemental Material (SM) \cite{SupplementalMaterial}, Sec. S4).

\begin{table}[t]
\caption{Constrained zero-temperature $^{20}$Ne shape endpoints: $D_0$ is the elongated reference and $D_-$ the more compact configuration. $Q_{20}$ is the undamped mass moment, $r_n$ the neutron rms radius and $\bar\Delta_n$ the mean neutron gap.}
\label{tab:anchors}
\begin{ruledtabular}
\begin{tabular}{lrrrr}
State & $E$ (MeV) & $Q_{20}$ (fm$^2$) & $r_n$ (fm) & $\bar\Delta_n$ (MeV)\\
$D_0$ & $-148.035$ & 151.30 & 3.146 & 1.585\\
$D_-$ & $-151.885$ & 104.82 & 3.012 & 1.519\\
\end{tabular}
\end{ruledtabular}
\end{table}
\subsection{Normal localization and spatial pairing measures}
For a representative time-reversed pair $i$, the thermal anomalous amplitude and quasiparticle energy are \cite{Goodman1981FT,Niu2013,Li2015FT}
\begin{align}
 \kappa_i&=\frac{\Delta_i^{\rm eff}}{2E_i}\tanh\frac{E_i}{2T},\nonumber\\
 E_i^2&=(\epsilon_i-\mu)^2+(\Delta_i^{\rm eff})^2.
 \label{eq:thermal}
\end{align}
The zero-temperature limit is understood. Here $\epsilon_i$ is the single-particle energy, $\mu$ fixes the mean particle number, and $\Delta_i^{\rm eff}$ is the cutoff-weighted gap. In the common gauge, $\Delta_i^{\rm eff}=-c_i\delta_i$, where $\delta_i$ is the signed attractive-field matrix element. Either member of the pair has normal occupation

\begin{equation}
 n_i=\frac12\left[1-\frac{\epsilon_i-\mu}{E_i}\tanh\frac{E_i}{2T}\right].
 \label{eq:normaloccupation}
\end{equation}
The same orbitals, weighted by $n_i$, construct the spin-resolved neutron density $\rho_{n\sigma}$, kinetic density $\tau_{n\sigma}$ and particle current $\mathbf j_{n\sigma}$. Their NLF is
\begin{align}
 C_{n\sigma}&=\left[1+(D_{n\sigma}/\tau_{n\sigma}^{\rm TF})^2\right]^{-1},\nonumber\\
 D_{n\sigma}&=\tau_{n\sigma}-\frac{|\nabla\rho_{n\sigma}|^2}{4\rho_{n\sigma}}
              -\frac{|\mathbf j_{n\sigma}|^2}{\rho_{n\sigma}},\nonumber\\
 \tau_{n\sigma}^{\rm TF}&=\frac35(6\pi^2)^{2/3}\rho_{n\sigma}^{5/3}.\label{eq:nlf}
\end{align}
Time reversal gives $\rho_{n\sigma}=\rho_n/2$; the spin-resolved particle currents cancel in the total current, but their squared terms are retained. The same normal-density diagnostic is evaluated with the fractional occupations of HF+BCS \cite{Reinhard2011,Bulgac2023}.

A deeply occupied orbital with $n_i\simeq1$ can contribute to the normal densities while carrying little pair amplitude. The NLF depends nonlinearly on the total normal densities and their gradients; near-Fermi anomalous reweighting can therefore alter the pairing tensor without a comparable change in the localization pattern. For the coordinate arguments 1 and 2, the cutoff-regulated scalar spin-singlet amplitude is

\begin{equation}
 \mathcal K(1,2)=\sum_i a_i\Phi_i(1,2),\qquad a_i=c_i\kappa_i.
 \label{eq:source}
\end{equation}
Writing $x=(\varrho,z)$, $\theta=\phi_1-\phi_2$ and $m_i=K_i-1/2$, its orbital kernel is
\begin{align}
 \Phi_i={}&U_i(x_1)U_i(x_2)\cos(m_i\theta)\nonumber\\
 &+D_i(x_1)D_i(x_2)\cos[(m_i+1)\theta].
 \label{eq:spinor}
\end{align}
$U_i,D_i$ are the real spin components of the nonrelativistic wavefunction, normalized with $d\mu(x)=2\pi\varrho\,d\varrho\,dz$. We use $d\Gamma=d\mu_1d\mu_2d\theta/\pi$, with $0\leq\theta\leq\pi$, and
\begin{equation}
 I_\kappa=\int\mathcal K^2\,d\Gamma.
 \label{eq:norm}
\end{equation}
For the cutoff-regulated antisymmetric neutron spin tensor $\widetilde\kappa^{\,c}$,
\begin{equation}
 \mathcal K(1,2)=\frac12\left[
 \widetilde\kappa^{\,c}_{\uparrow\downarrow}(1,2)
 -\widetilde\kappa^{\,c}_{\downarrow\uparrow}(1,2)\right].
 \label{eq:singlet}
\end{equation}
The antisymmetric spin combination is the normalized singlet component divided by $\sqrt2$; its coincident value is the local neutron source $S_n$. Each Kramers representative is counted once. The dimensionless $I_\kappa$ measures anomalous strength, separately from the normal direct and exchange sectors. Both species enter the self-consistent fields and density-dependent pairing rearrangement; the Ne spatial pairing analysis concerns neutrons. The SM, Secs. S1--S2, specifies the functional, normalization, cutoff and cross-state orbital matching.

The pair-center and relative radii are $R=|\mathbf r_1+\mathbf r_2|/2$ and $r=|\mathbf r_1-\mathbf r_2|$, measured from the nucleon center of mass. Projecting $\mathcal K^2d\Gamma/I_\kappa$ onto these radii defines the normalized anomalous-strength distribution $p(R,r)$. The conditional distribution is
\begin{equation}
 \int p\,dR\,dr=1,\qquad
 W(R,r)=\frac{p(R,r)}{\int p(R,r')\,dr'}.
 \label{eq:maps}
\end{equation}
The units are fm$^{-2}$ and fm$^{-1}$. Conservative cell masses include the angular measure, radial Jacobians and overflow support. The finite cosine expansion is integrated analytically over each cell's angular intervals; the regular radial bins are 0.2 fm wide. Display smoothing never changes these masses. For cell masses $P_{ab}$, row masses $u_a=\sum_bP_{ab}$ and conditional masses $w_{b|a}=P_{ab}/u_a$, the reported distances are
\begin{align}
 D_p&=\tfrac12\sum_{ab}|P^{(t)}_{ab}-P^{(0)}_{ab}|,\nonumber\\
 D_W&=\tfrac12\sum_a u^{(0)}_a\sum_b|w^{(t)}_{b|a}-w^{(0)}_{b|a}|.
 \label{eq:TV}
\end{align}
They use full support, including overflow bins; zero-mass rows have zero conditional arrays. The relative row threshold used to display $W$ removes no mass from these metrics. Normalized pairing maps are evaluated only for appreciable anomalous sources (SM Sec. S4). Radial maps average over orientation, so small $R$ alone does not identify a pair spanning the ends.

To resolve connections between the NLF lobes, define fixed end regions L ($z<-2.2$ fm) and R ($z>2.2$ fm). Their planes are the outer cell faces adjacent to the density-weighted radial NLF valley of $D_0$, selected before comparing states; separate boundary shifts test this regional choice (SM Sec. S5). Same-end and opposite-end sectors, $S=LL\cup RR$ and $X=LR\cup RL$, have absolute integrals and fractions

\begin{equation}
 M_\Omega=\int_\Omega\mathcal K^2d\Gamma,
 \qquad F_\Omega=M_\Omega/I_\kappa.
 \label{eq:regions}
\end{equation}
Configurations with at least one central endpoint form the remainder $1-F_S-F_X$. A ``pairing connection'' is the two-endpoint anomalous integral on the NLF-defined background; an enhanced opposite-end connection has larger absolute $M_X$. The fixed planes select the oriented subset $r\geq|z_1-z_2|\geq4.4$ fm, excluding central endpoints and short-separation pairs and accounting for its modest norm fraction. The pair-center-dependent coherence length instead measures rms relative separation \cite{Pillet2007,Pillet2010,Lasseri2018}.

\subsection{Orbital decomposition and response criteria}\label{sec:criterion}
For the complete set of orbital kernels, let $G_{ij}=\int\Phi_i\Phi_jd\Gamma$ and $G_{\Omega,ij}=\int_\Omega\Phi_i\Phi_jd\Gamma$. Then
\begin{equation}
 I_\kappa=a^{\mathsf T}Ga,\qquad M_\Omega=a^{\mathsf T}G_\Omega a.
 \label{eq:gram}
\end{equation}
The complete Ne neutron budget contains 24 self-terms and 276 distinct interference terms, each cross term including both orderings. The full pool is retained in the tracked HF representation (SM Sec. S2 and accompanying data).

Define the mean end-region strength and regional contrast,
\begin{equation}
 \mathcal C=\tfrac12(M_S+M_X),\qquad
 \mathcal O=\tfrac12(M_S-M_X).
 \label{eq:CO}
\end{equation}
Hence $M_X=\mathcal C-\mathcal O$ and $M_S=\mathcal C+\mathcal O$. For definite-parity orbitals with eigenvalues $\pi_i$, inversion of one endpoint multiplies $\Phi_i$ by $\pi_i$. If this transformation maps the full S domain onto X and preserves the measure, $G_{X,ij}=\pi_i\pi_jG_{S,ij}$. Opposite-parity interference then contributes to $\mathcal O$, while self and same-parity terms contribute to $\mathcal C$. Equation~(\ref{eq:CO}) remains exact without that parity identification.

For a fixed physical region $\Omega$, the local response of its fraction follows from Eq.~(\ref{eq:gram}):
\begin{align}
 \delta F_\Omega={}&\sum_i L_{\Omega i}\,\delta\ln|a_i|
 +\frac{a^{\mathsf T}(\delta G_\Omega-F_\Omega\delta G)a}{I_\kappa},
 \nonumber\\
 L_{\Omega i}={}&\frac{2a_i[(G_\Omega a)_i-F_\Omega(Ga)_i]}{I_\kappa},
 \qquad\sum_i L_{\Omega i}=0.
 \label{eq:sensitivity}
\end{align}
For nonzero coefficients, $L_{\Omega i}$ measures the interference-inclusive sensitivity to relative anomalous reweighting. Common coefficient scaling cancels from the first term; orbital-dependent changes contribute through their signed sensitivities. The kernel term captures reinforcing or opposing spatial feedback. Finite changes retain the complete endpoint sums.

Since $M_\Omega=F_\Omega I_\kappa$ by definition, absolute enhancement for two states with $0<I_1<I_0$ and $F_{\Omega,0}>0$ is equivalent to
\begin{equation}
 M_{\Omega,1}>M_{\Omega,0}
 \quad\Longleftrightarrow\quad
 \frac{F_{\Omega,1}}{F_{\Omega,0}}>\frac{I_0}{I_1}.
 \label{eq:absolutecondition}
\end{equation}
Here $I_s=I_\kappa(s)$. Equation~(\ref{eq:absolutecondition}) expresses the exact normalization balance; Eq.~(\ref{eq:sensitivity}) and the finite endpoint sums resolve the weight and kernel changes that produce it. Moving the region adds its boundary contribution to $\delta G_\Omega$.

For reference coefficients $a$ and matched target coefficients $b$, the departure from proportional attenuation is measured over the complete pool in the reference Gram metric:
\begin{equation}
 \alpha=\frac{a^{\mathsf T}G_0b}{a^{\mathsf T}G_0a},\qquad
 \eta=\frac{(b-\alpha a)^{\mathsf T}G_0(b-\alpha a)}{b^{\mathsf T}G_0b}.
 \label{eq:common}
\end{equation}
Here $\alpha$ is the common multiplier and $\eta$ the remaining norm fraction. The amplitude $\sqrt{\eta}$ measures nonproportional orbital participation, not numerical error. If the spatial kernels also change little, $\mathcal K_1\simeq\alpha\mathcal K_0$: absolute integrals scale approximately as $\alpha^2$, while normalized distributions nearly retain their shape.

\section{Results and discussion}\label{sec:results}
\subsection{Deformation-induced pairing redistribution in \texorpdfstring{$^{20}$Ne}{20Ne}}
\begin{figure*}[!t]
\centering\includegraphics[width=\textwidth]{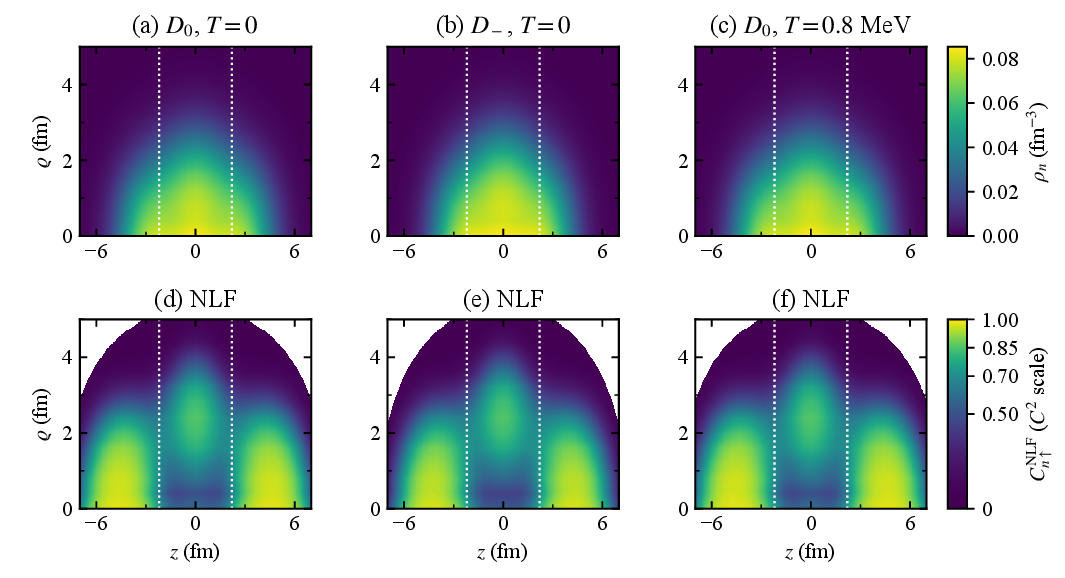}
\caption{Neutron density (top) and NLF (bottom) for the constrained SLy5 $^{20}$Ne shape and thermal states. Dotted planes selected from the reference NLF valley define fixed end regions. Scales are common within each row. The NLF palette maps $C$ to $C^2$ over 0--1; the colorbar labels actual $C$. Values at $\rho_n<10^{-4}$ fm$^{-3}$ are masked. Continuous contours use display interpolation only; integration data are unchanged.}
\label{fig:background}
\end{figure*}

\begin{figure*}[!t]
\centering\includegraphics[width=\textwidth]{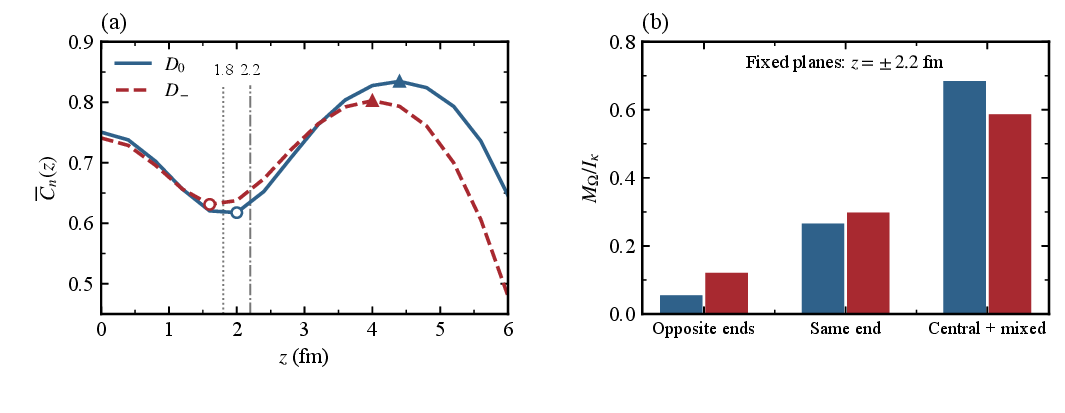}
\caption{Localization regions and their share of the anomalous norm in cold $^{20}$Ne. (a) Radial density-weighted NLF profiles on the positive $z$ half-axis. Circles mark the intervening valleys and triangles the end maxima. The 2.2 fm plane is used for both states; 1.8 fm is the alternative plane transported with the compact-state valley. The opposite-end domain includes the end lobes. (b) Fractions of the full anomalous norm at the common 2.2 fm boundary. The remainder contains all configurations with at least one central endpoint. Absolute integrals are given in Table~\ref{tab:core}; the profile definition is given in SM Sec. S5.}
\label{fig:regions}
\end{figure*}

The cold $D_0$ and $D_-$ configurations compare the elongated and compact slices, while the $D_0$ thermal states retain the reference quadrupole target (Table~\ref{tab:core}). Recognizable NLF end lobes persist under contraction, with maxima beyond the fixed planes and valleys near the boundaries [Figs.~\ref{fig:background} and \ref{fig:regions}]. The fixed-end integral $M_X$ isolates the anomalous strength joining these lobes, excluding configurations with a central endpoint. On this common spatial reference, $D_-$ has a smaller total anomalous norm, mean gap and pairing-energy magnitude, yet more than twice the absolute opposite-end strength. The growth of its regional fraction outweighs the modest loss of total norm. Both end-region integrals increase while the still-dominant central and mixed contribution decreases. These are changes in a nonconserved anomalous strength.

\begin{table*}[t]
\caption{Internal energy and neutron pairing under deformation and heating in SLy5 $^{20}$Ne. $E$ includes the full center-of-mass correction after variation and excludes the constraint term; at finite temperature it is the internal energy, not the free energy. The anomalous norm and absolute regional integrals are dimensionless. $M_S$ joins points in the same end region, $M_X$ joins opposite ends, and $F_X=M_X/I_\kappa$.}
\label{tab:core}
\begin{ruledtabular}
\begin{tabular}{lrrrrrrr}
State & $T$ (MeV) & $E$ (MeV) & $E_{{\rm pair},n}$ (MeV) & $I_\kappa$ & $M_S$ & $M_X$ & $F_X$ (\%)\\
$D_0$ & 0.0 & -148.035 & -2.011 & 0.2160 & 0.0570 & 0.0115 & 5.32\\
$D_0$ & 0.4 & -148.020 & -1.991 & 0.2138 & 0.0565 & 0.0114 & 5.32\\
$D_0$ & 0.8 & -147.307 & -1.285 & 0.1378 & 0.0369 & 0.0075 & 5.42\\
$D_-$ & 0.0 & -151.885 & -1.753 & 0.2046 & 0.0606 & 0.0243 & 11.88\\
\end{tabular}
\end{ruledtabular}
\end{table*}

\begin{figure*}[!t]
\centering\includegraphics[width=\textwidth]{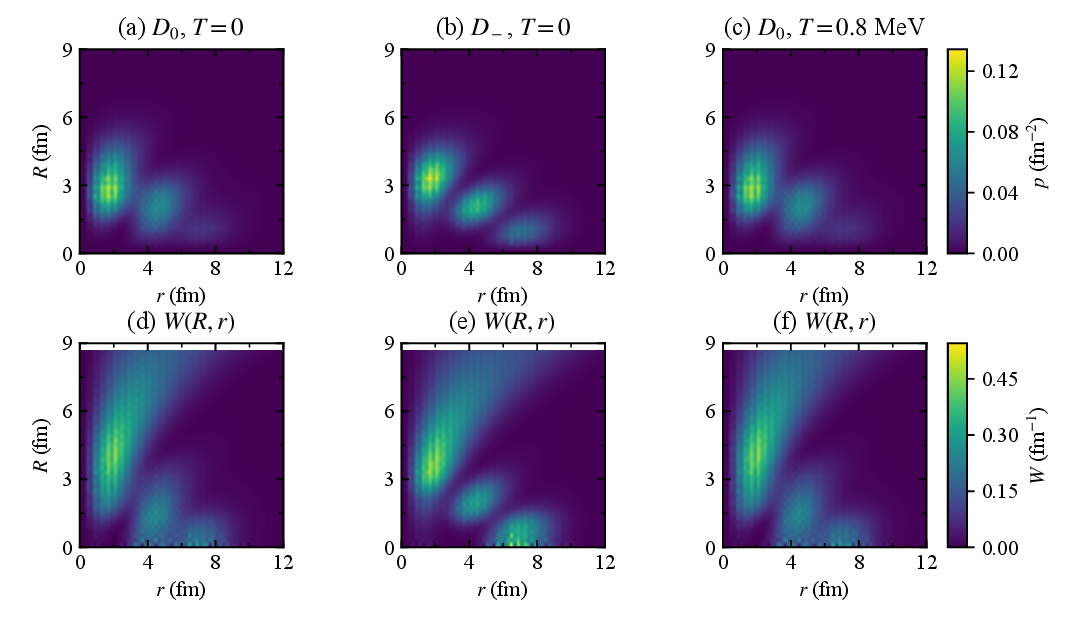}
\caption{Global $p(R,r)$ (top) and conditional $W(R,r)$ (bottom), in the column order of Fig.~\ref{fig:background}. Contraction redistributes the pair distribution; heating mainly reduces its absolute strength. The radial maps average orientations, while the L/R integrals resolve the endpoints. Colors are common within each row. The display is cropped from full-support distributions; $W$ suppresses rows below $10^{-4}$ of the maximum row density. Smoothing preserves the bin masses.}
\label{fig:maps}
\end{figure*}

\begin{figure*}[!t]
\centering\includegraphics[width=\textwidth]{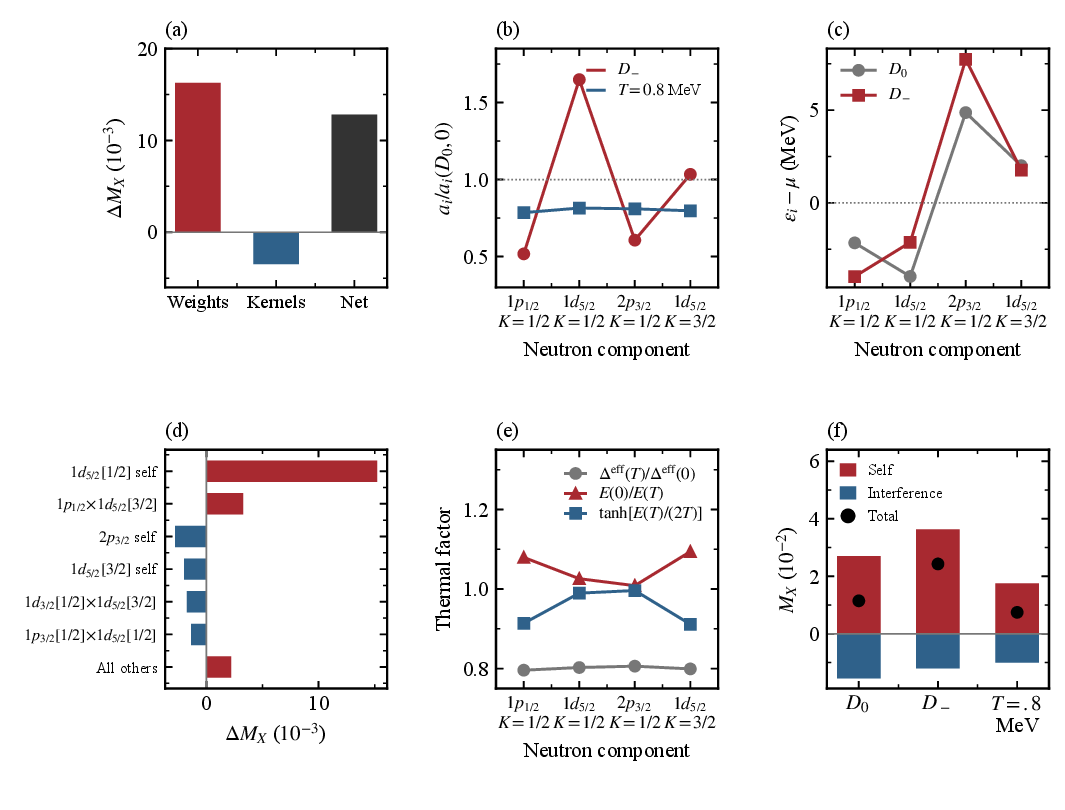}
\caption{Complete neutron-orbital response of the $^{20}$Ne pairing connections. (a) Complete opposite-end change split into anomalous-weight and spatial-kernel allocations. (b) Coefficient ratios $a_i/a_i(D_0,0)$ for the labelled $nlj,K$ components; the $2p_{3/2}$ component retains its $1f_{7/2}$ admixture (Table~\ref{tab:parentage}). (c) Single-particle energies relative to $\mu$. (d) Signed changes of $M_X$, with $K$ in square brackets; ``All others'' completes the 300-term sum. (e) Thermal factors in Eq.~(\ref{eq:factors}). (f) Complete self and interference sums. Lines connect discrete components. The mesh comparison is given in SM Sec. S5.}
\label{fig:mechanism}
\end{figure*}

The orientation-averaged $p$ and $W$ maps resolve the accompanying changes of pair-center and separation structure (Fig.~\ref{fig:maps}); the endpoint integrals identify the part joining the two lobes. Within the matched HF representation, changing the kernels alone does not explain the gain: compact-state kernels with reference weights lower $M_X$, whereas compact-state weights on reference kernels raise it above its final value [Fig.~\ref{fig:mechanism}(a)]. These endpoint substitutions separate contributions without defining additional self-consistent states. Contraction changes the spectrum, gaps and spatial support together; the question is why the resulting reweighting favors the opposite-end connection.

Eight orbitals, matched by wavefunction overlap, account for nearly all of the cold opposite-end gain. Their 36 terms and the remaining 264 enter every total (SM Sec. S2 and accompanying data). At zero temperature, Eq.~(\ref{eq:thermal}) makes anomalous participation depend on the distance from $\mu$ relative to the effective gap, $|\epsilon_i-\mu|/|\Delta_i^{\rm eff}|$, with the cutoff fixed. Contraction moves the predominantly $1p_{1/2}(K=1/2)$ orbital farther below $\mu$ and reduces its gap magnitude. Both changes suppress its anomalous weight. The mixed $1d_{5/2}/2s_{1/2}(K=1/2)$ orbital moves closer to $\mu$ and its gap magnitude increases, reinforcing the opposite response [Fig.~\ref{fig:mechanism}(b,c); SM Sec. S2]. Their energy ordering reverses while both remain below $\mu$.

These changes distinguish normal occupation from anomalous participation. The $1p_{1/2}$ orbital becomes more fully occupied while losing pair amplitude; the mixed $sd$ orbital becomes less fully occupied while gaining it. Both continue to contribute to the normal density, so their reweighting need not remove the NLF end lobes. Angular projections and radial-node counts identify the dominant $nlj$ components of the deformed $K$ branches (Table~\ref{tab:parentage}; SM Sec. S7). Squared cross-state overlaps above 0.97 track the four displayed orbitals, including the explicitly mixed $2p_{3/2}/1f_{7/2}$ component.

\begin{table}[t]
\caption{Neutron $nlj$ component labels and selected weights (integer percentages) in the cold states. Here $n=1,2,\ldots$ counts radial nodes plus one; $K^\pi$ identifies the deformed branch. Smaller components remain in the complete budget. Radial projections and multiplet assignments are given in SM Sec. S7.}
\label{tab:parentage}
\begin{ruledtabular}
\begin{tabular}{cccrr}
Channel & $K^\pi$ & Component & $D_0$ & $D_-$\\
$1p_{1/2}$ & $1/2^-$ & $1p_{1/2}$ & 78 & 84\\
$1d_{5/2}$ & $1/2^+$ & $1d_{5/2}$ & 55 & 65\\
   &           & $2s_{1/2}$ & 29 & 25\\
$2p_{3/2}/1f_{7/2}$ & $1/2^-$ & $2p_{3/2}$ & 45 & 47\\
   &           & $1f_{7/2}$ & 33 & 37\\
$1d_{5/2}$ & $3/2^+$ & $1d_{5/2}$ & 85 & 91\\
\end{tabular}
\end{ruledtabular}
\end{table}

The regional effect of this reweighting depends on orbital spatial support and interference. The mixed $1d_{5/2}/2s_{1/2}(K=1/2)$ orbital supplies the largest $M_X$ gain: its increased anomalous weight outweighs a weaker opposite-end self-kernel [Fig.~\ref{fig:mechanism}(d)]. Its same-end contribution also grows, raising $\mathcal C$. In contrast, the $1p_{1/2}(K=1/2)\times1d_{5/2}(K=3/2)$ interference is constructive within an end and destructive between ends. The reduced $1p_{1/2}$ weight and cross-kernel magnitude weaken both contributions. A loss of pair participation can therefore strengthen the opposite-end connection by reducing cancellation, even as it weakens a constructive same-end term. This contrast loss reinforces $M_X=\mathcal C-\mathcal O$ but partly offsets the growth of $M_S=\mathcal C+\mathcal O$.

The leading self-term gain exceeds the complete increase. Losses from the mixed $2p_{3/2}/1f_{7/2}$ and $1d_{5/2}(K=3/2)$ self-terms compensate part of it. The former loses anomalous weight; the latter has little weight change but a smaller opposite-end kernel despite a larger $1d_{5/2}$ fraction. Angular purity alone therefore does not determine regional support. Interference shapes the contrast but is not required for the entire enhancement; Fig.~\ref{fig:mechanism}(d) retains every remaining term.

Averaging the two substitution orders assigns a positive opposite-end change to the weights and a net negative change to the kernels [Fig.~\ref{fig:mechanism}(a)]. Although the reduced destructive $1p_{1/2}\times1d_{5/2}(K=3/2)$ cross-kernel contributes positively, losses elsewhere dominate the kernel sum. The weight gain exceeds this opposing spatial response. Each crossed fraction uses its own full-support denominator; the absolute numerator gain dominates the increase of $F_X$ (SM Secs. S3 and S6).

The same terms explain how this regional gain coexists with a lower total norm. Depleted $1p_{1/2}$ participation reduces its positive full-space self-term and its constructive interference with the $1p_{3/2}(K=3/2)$ component, while releasing destructive opposite-end interference with $1d_{5/2}(K=3/2)$. This opposite-parity cross term nearly cancels over full space but contributes with opposite signs within and between the ends. The complete global budget instead gives nearly canceling weight gains and losses, with a larger negative kernel allocation (SM Sec. S6). The global and regional balances therefore differ even though they contain exactly the same orbital pool.

The distinct roles of these orbital contributions are collected in Table~\ref{tab:roles}. They refer to the present contraction and fixed end regions; a single orbital can enter several roles through its self and interference terms.

\begin{table*}[t]
\caption{Orbital and interference contributions to the cold $D_0\to D_-$ response. The roles are nonexclusive and refer to the fixed end regions and matched HF representation. Component labels retain their deformed mixing; the complete numerical budget is given in SM Sec. S6.}
\label{tab:roles}
\begin{ruledtabular}
\begin{tabular}{lll}
\parbox[t]{0.28\textwidth}{\raggedright Contribution} &
\parbox[t]{0.30\textwidth}{\raggedright Dominant response} &
\parbox[t]{0.36\textwidth}{\raggedright Physical role}\\
\hline
\parbox[t]{0.28\textwidth}{\raggedright Mixed $1d_{5/2}/2s_{1/2}$, $K=1/2$, self-term} &
\parbox[t]{0.30\textwidth}{\raggedright Weight gain outweighs the opposing kernel response.} &
\parbox[t]{0.36\textwidth}{\raggedright Principal positive contribution to $M_X$; also raises $I_\kappa$.}\\[3pt]
\parbox[t]{0.28\textwidth}{\raggedright $1p_{1/2}(K=1/2)$ self-term and interference with $1p_{3/2}(K=3/2)$} &
\parbox[t]{0.30\textwidth}{\raggedright Reduced $1p_{1/2}$ participation weakens positive self and same-parity terms.} &
\parbox[t]{0.36\textwidth}{\raggedright Removes global positive strength while the same orbital also regulates regional cancellation.}\\[3pt]
\parbox[t]{0.28\textwidth}{\raggedright $1p_{1/2}(K=1/2)\times1d_{5/2}(K=3/2)$} &
\parbox[t]{0.30\textwidth}{\raggedright Destructive opposite-end interference becomes weaker.} &
\parbox[t]{0.36\textwidth}{\raggedright Adds to $M_X$ while reducing constructive same-end interference; the full-space term nearly cancels by parity.}\\[3pt]
\parbox[t]{0.28\textwidth}{\raggedright $1d_{5/2}$, $K=3/2$, self-term} &
\parbox[t]{0.30\textwidth}{\raggedright Kernel loss exceeds the small positive weight contribution.} &
\parbox[t]{0.36\textwidth}{\raggedright Reduces $M_X$ despite a large reference sensitivity and little change in anomalous weight.}\\[3pt]
\parbox[t]{0.28\textwidth}{\raggedright Mixed $2p_{3/2}/1f_{7/2}$, $K=1/2$, self-term} &
\parbox[t]{0.30\textwidth}{\raggedright Anomalous weight decreases.} &
\parbox[t]{0.36\textwidth}{\raggedright Offsets part of the opposite-end gain.}\\
\end{tabular}
\end{ruledtabular}
\end{table*}

The full pool retains cutoff-edge terms, including $1d_{3/2}(K=1/2)\times1d_{5/2}(K=3/2)$. The well-matched mixed $p/f$ level approaches the instantaneous window edge while retaining its frozen cutoff. Replacing this cutoff at fixed occupations and wavefunctions further increases $M_X$ by reducing destructive interference (SM Sec. S2); changing the window self-consistently would also modify the coupled state. The full matching table retains the weak contributions and exchanged orbital assignments.

One-cell shifts of the fixed planes give $M_X(D_-)/M_X(D_0)=2.11$--2.14. Transporting the compact-state planes with the NLF valley from 2.2 to 1.8 fm gives 2.66, including the changed endpoint geometry. All tested regions retain the enhancement, growing $1d_{5/2}(K=1/2)$ self-term and weaker destructive $1p_{1/2}\times1d_{5/2}(K=3/2)$ interference. Geometry affects the magnitude and subleading allocations: the $1d_{5/2}(K=3/2)$ self-term change reverses sign with the transported boundary.

\subsection{Thermal response of pairing in \texorpdfstring{$^{20}$Ne}{20Ne}}
Heating at the $D_0$ quadrupole constraint nearly preserves both the density-supported NLF and normalized pairing patterns (Figs.~\ref{fig:background} and \ref{fig:maps}). At 0.8 MeV the total anomalous norm falls by roughly one third, while $p$ and $W$ change much less than under contraction. The weak 0.4 MeV response supplies an intermediate point. The smaller thermal change of the NLF also persists in the common density-weighted metric (SM Sec. S4).

Heating supplies a direct counterexample to interpreting a larger normalized fraction as a stronger connection. At 0.8 MeV, $F_X$ increases by about 2\%, whereas the absolute $M_X$ decreases by about 35\%: the full anomalous norm falls faster than the regional strength. The fractional gain is therefore insufficient to satisfy Eq.~(\ref{eq:absolutecondition}). In the complete orbital budget, reduced destructive interference is outweighed by the declining positive self-strength [Fig.~\ref{fig:mechanism}(f)].

Heating tests whether the same regionally sensitive orbitals are selectively reweighted. The explicit thermal factor is orbital dependent, while the gaps and quasiparticle energies respond self-consistently. With the explicit cutoff fixed, Eq.~(\ref{eq:thermal}) gives
\begin{equation}
 \frac{a_i(T)}{a_i(0)}=
 \frac{\Delta_i^{\rm eff}(T)}{\Delta_i^{\rm eff}(0)}
 \frac{E_i(0)}{E_i(T)}\tanh\frac{E_i(T)}{2T}.
 \label{eq:factors}
\end{equation}
The three factors share the same self-consistent quasiparticle energies and describe coupled changes. The thermal factor $\tanh[E_i(T)/(2T)]$ is smaller for the near-Fermi $1p_{1/2}$ and $1d_{5/2}(K=3/2)$ components than for the $1d_{5/2}(K=1/2)$ and mixed $2p_{3/2}/1f_{7/2}$ components. The former also undergo a larger quasiparticle-energy decrease, which strengthens the compensating inverse-energy factor [Fig.~\ref{fig:mechanism}(e)]. Their similar gap reductions then bring the net coefficient ratios close to a common multiplier of 0.8.

At 0.8 MeV, Eq.~(\ref{eq:common}) gives $\alpha=0.799$ and a noncommon amplitude $\sqrt{\eta}\simeq1.1\%$, compared with about 39\% under deformation. Together with weak kernel changes and nearly unit dominant-orbital overlaps, this gives $\mathcal K(T)\simeq\alpha\mathcal K(0)$. Positive self-terms and signed interference then decrease by nearly the same factor $\alpha^2$, preserving their relative balance. The connection loses strength without the selective release of cancellation seen under contraction; constructive same-end and destructive opposite-end interference remain in the complete budget.

The reference sensitivities distinguish susceptibility from realized response (SM Sec. S3.1). Under contraction, the $1d_{5/2}(K=3/2)$ branch has a larger positive sensitivity than the mixed $sd$ branch, but its weaker relative reweighting produces a smaller positive contribution to the linear change of $F_X$. Its declining $M_X$ self-term further shows how kernel loss can overcome a favorable weight change. The response thus depends on the perturbation of the anomalous weights as well as the signed regional sensitivities and spatial feedback. Nearly proportional thermal changes instead give small signed sensitivity-weight sums. The local relation reproduces the fixed-kernel thermal responses at 0.4 and 0.8 MeV within 1\%, but underestimates the fixed-kernel contraction response by about 28\%, requiring nonlinear reweighting. The opposing kernel term remains appreciable relative to the small net change even at 0.4 MeV. Common attenuation follows here from the coupled gap and quasiparticle compensation, not from temperature alone; other subshell structures and continuum coupling can yield pairing persistence or reentrance \cite{Li2015FT}.

\subsection{Pairing quenching and localization in \texorpdfstring{$^{144}$Ba}{144Ba}}
Octupole-deformed Ba tests the same distinction when reflection symmetry is broken and the quadrupole and octupole moments respond freely. Opposite-parity levels near the Fermi surface favor octupole deformation \cite{Agbemava2016prc}; tensor-induced spin--orbit changes can alter these levels in Ba \cite{Xue2026Ba}. In the present fixed-interaction comparison, heating changes normal occupations and anomalous weights together. The questions are whether this shape freedom produces selective redistribution and whether the normal spatial structure survives when the anomalous connection is quenched. At 0.6 MeV the neutron source remains resolved, whereas the proton anomalous source is numerically unresolved and has no meaningful normalized pairing map.

\begin{figure*}[!t]
\centering\includegraphics[width=\textwidth]{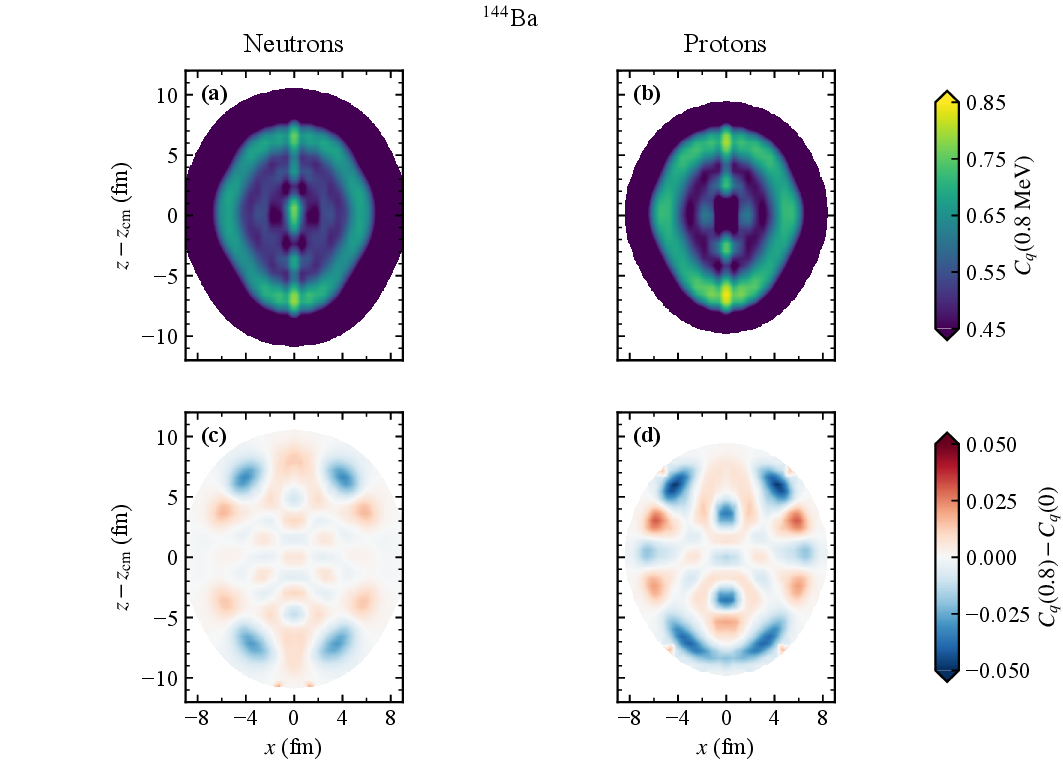}
\caption{Normal localization after thermal quenching of the anomalous sources in $^{144}$Ba. Neutrons are on the left and protons on the right. Upper panels show the NLF at 0.8 MeV in the 24 fm box; lower panels show its pointwise change from the cold reference, on a common symmetric color scale. Coordinates are centered on each state's total-nucleon center of mass. Values with $\rho_q<10^{-4}$ fm$^{-3}$ are blank; differences use the common supported region. The hot state retains the cold cutoff. Smooth display interpolation leaves the integration data unchanged; local differences are distinct from the density-weighted rms measure in SM Sec. S4.}
\label{fig:ba_nlf}
\end{figure*}

At 0.6 MeV in the 24 fm box, both the neutron norm and opposite-half-space integral $M_H$ retain about 36\% of their cold strength. The Ba half-spaces meet at the center of mass, so their absolute integrals are not compared with those between the separated Ne end regions. The complete 155-orbital budget assigns almost the entire $M_H$ loss to attenuation of the anomalous amplitudes, with negligible kernel response. Positive self-terms and net constructive interference decline together; a common multiplier near 0.60 leaves a roughly 5\% noncommon amplitude in the cold Gram metric (SM Sec. S6).

\begin{figure*}[!t]
\centering\includegraphics[width=\textwidth]{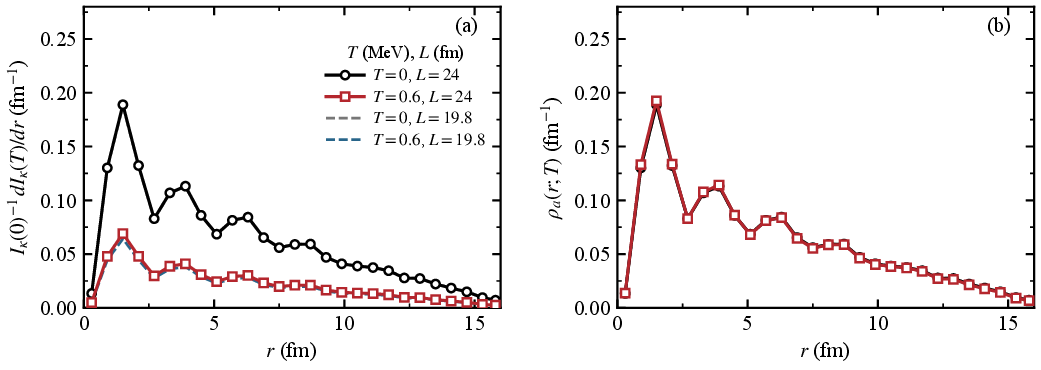}
\caption{Thermal attenuation and spatial shape of the neutron anomalous strength in octupole-deformed $^{144}$Ba at $T=0$ and 0.6 MeV. Solid and dashed curves denote box extents $L=24$ and 19.8 fm, respectively; the legend applies to both panels. (a) The relative-distance distribution $I_\kappa(0)^{-1}\,dI_\kappa(T)/dr$, normalized to each box's cold total norm, retains the thermal loss of strength. (b) The unit-normalized distribution $\rho_a(r;T)=I_\kappa(T)^{-1}\,dI_\kappa(T)/dr$ shows the nearly preserved spatial shape. The distributions integrate over the pair-center radius $R$. For display, adjacent 0.2 fm bins are conservatively combined into 0.6 fm bins (0.4 fm for the terminal bin), preserving the integrated strengths. Symbols mark every displayed bin center on the solid curves. The displayed range is $r\leq16$ fm; normalization includes the full support and overflow.}
\label{fig:ba_distance}
\end{figure*}

The full-support radial distributions show little redistribution in pair-center and separation coordinates (Fig.~\ref{fig:ba_distance}). In the 24 fm box, $p(R,r)$ changes by about 1.4\% in total variation, with a small shift toward shorter separation and comparable attenuation in the windows $r\leq3$ fm, $3<r<6$ fm and $r\geq6$ fm. The 19.8 fm comparison retains this weak shape response. Conditional $W$ also changes little, using supported mean-row weights for Ba rather than the Ne reference-row metric (SM Sec. S4).

At 0.8 MeV, negligible neutron and proton anomalous sources coexist with a density-supported interior NLF and nonzero octupole moment (Fig.~\ref{fig:ba_nlf}). The octupole magnitude first rises slightly, then falls below its cold value; NLF rms changes stay below 0.01 for both species at both temperatures (SM Sec. S4). The normal occupations can sustain structured densities and localization after the anomalous amplitudes reach the numerical normal limit. A surviving NLF pattern therefore does not imply a surviving pairing connection. Normalized pairing maps remain undefined for these unresolved sources. Positive-energy neutron occupations produce a dilute exterior, restricting the comparison to finite-box interior patterns; continuum coupling can modify spatial pairing in weakly bound systems \cite{Dobaczewski1996}.

\subsection{Discussion}\label{sec:discussion}
The two perturbations distinguish selective participation from common attenuation on an identifiable localization background. At zero temperature and fixed cutoff, an orbital's distance from the chemical potential relative to its self-consistent gap controls its anomalous participation. Its regional effect depends on spatial support and interference with the rest of the pool. In Ne, contraction favors the mixed $sd$ component's positive strength and weakens destructive cross terms through depleted $p$ participation. This weight gain exceeds the opposing spatial feedback. Heating preserves nearly the same relative participation through compensation of thermal and inverse-energy factors together with similar gap reductions, so it mainly changes the overall strength.

For another configuration, the relevant question is whether the perturbation selectively changes orbitals with substantial signed regional sensitivities. Large weight changes in insensitive channels, or compensating changes across channels, need not redistribute the regional fraction. Even a favorable fractional response must exceed the total-norm loss to give absolute enhancement, and changing kernels can reverse a weight-only inference. Nearly proportional coefficients preserve the pattern if the spatial kernels also remain stable; proportionality alone does not ensure it. The NLF establishes whether the regions themselves remain identifiable; it does not assign the sign or magnitude of their anomalous connection.

Parity interference and proportional-weight cancellation are established features of spatial pairing \cite{Pillet2010,Lasseri2018}. The present comparison connects them to a self-consistent spectrum--gap response and shows how absolute regional enhancement survives opposing spatial feedback. It supplies orbital conditions to examine, rather than a universal enhancement rule based on $nlj$ labels or contraction alone. For example, NLF-identified prefragments before scission \cite{Zhang2016Localization} provide regions in which the same separation of participation and spatial feedback could be applied along a prescribed path. Such static strengths characterize pair amplitudes, not currents or dynamical transfer rates.

Electromagnetic moments constrain the normal intrinsic background on which pairing connections are defined. For $Q_{0p}=e\int(2z^2-\varrho^2)\rho_p\,d^3r$, the axial $K=0$ rotor limit gives $B(E2;0\to2)=5Q_{0p}^2/(16\pi)$ and $Q_s(2)=-2Q_{0p}/7$. The conditional $D_0$ conversions approach the measured $Q_s$ but overestimate the transition strength \cite{Mehl2025,Xue2025} (SM Sec. S5). Shape softness and rotational overlaps limit this conversion \cite{Robledo2012}; angular-momentum projection and configuration mixing connect intrinsic configurations to measured spectra and matrix elements \cite{Xue2024prc,Xue2024prc2,Xue2025}. The $^{144,146}$Ba octupole measurements constrain the cold collective background \cite{Bucher2016PRL,Bucher2017PRL}. Collective and cranking calculations address its spectra, transitions and persistence under rotation \cite{Robledo2011prc,Nomura2015prc,LiZK2024NST}, separately from the thermal connection response considered here.

Two-neutron transfer probes a number-changing amplitude,
\begin{equation}
 T_\nu(1,2)=\langle\Psi_{N-2,\nu}|\psi(2)\psi(1)|\Psi_N\rangle,
\end{equation}
folded with reaction dynamics \cite{Potel2013,Scamps2013}. Distinct initial and final vacua and particle-number treatment enter this connection \cite{Grasso2012Transfer}. Our same-state source identifies spatial components that may contribute to a transfer form factor, but is not a cross section: a fixed-number state can have zero same-state anomalous expectation and finite transfer amplitude. A laboratory strength requires a separate state-to-state calculation. Low-energy isoscalar response connects cluster motion to normal transition densities \cite{Mercier2021prc}, without directly measuring $M_X$.

The specified paired intrinsic branches have undetermined energy ordering against competing normal configurations. Their fixed mean particle numbers coexist with broken-symmetry and thermal variances, distinct from the regulated singlet norm (SM Sec. S8). Absolute strengths depend on the functional, cutoff and center-of-mass prescription. Proton--neutron pairing, symmetry restoration and configuration mixing require extensions of this HF+BCS description \cite{Frauendorf2014,DuguetRyssens2020}. With different spatial backgrounds and center-of-mass treatments, Ne and Ba test distinct structural settings rather than a controlled mass variation.

Refining the Ne mesh from 0.4 to 0.3 fm preserves the opposite-end gain, leading orbital contributions and negative kernel allocation; the smaller total-norm decrease is more mesh sensitive (SM Sec. S5). The cold Ba comparison from 24 to 30 fm changes both neutron integrals by less than 0.3\%. Both thermal boxes retain weight-dominated attenuation, but their separate cold states, cutoffs and continuation paths prevent attributing quantitative differences to volume alone. These comparisons test the mechanism at the stated resolutions and volumes; they do not provide a continuum extrapolation or a box-independent critical temperature.

\section{Summary and conclusions}\label{sec:conclusions}
The joint NLF and pairing-tensor analysis shows how a connection between localized nuclear regions can strengthen during overall pairing loss. Along the constrained $^{20}$Ne contraction, the coupled spectrum and gap response increases anomalous participation of the mixed $1d_{5/2}/2s_{1/2}$ component while reducing that of a predominantly $1p_{1/2}$ component. Increased positive strength and weaker destructive interference raise the absolute opposite-end integral. Complete orbital substitutions show that this gain overcomes a net opposing change in spatial kernels, a balance preserved by mesh refinement.

Heating at the reference quadrupole constraint gives a different balance. Thermal suppression and inverse-quasiparticle-energy changes partly compensate, and similar gap reductions keep the dominant coefficients close to a common multiplier. Absolute strength decreases while the relative self and interference contributions, and hence the spatial pattern, are nearly retained. Octupole-deformed $^{144}$Ba also exhibits weight-dominated attenuation on a freely responding shape background. At higher temperature, normal localization persists after the anomalous sources become negligible.

Selective participation strengthens a regional connection when its net weight contribution to the complete absolute integral exceeds an opposing kernel contribution. Nearly common participation changes preserve the pattern when the kernels remain stable. These conditions link spectral and pairing changes to regional responses in other configurations with identifiable localization. Absolute integrals are essential to distinguish an enhanced connection from a larger normalized fraction; the persistence of the normal localization background alone determines neither.

\section*{Acknowledgements}
This work was supported by the National Natural Science Foundation of China under 
Grants No. 12505137, 12575124, and the Key Scientific Research Projects of Universities in Henan Province (Grant No. 26B140011).

\section*{Data availability}
The data supporting the findings of this study are available from the author upon reasonable request.

\bibliography{references}
\end{document}